\documentstyle{ar_prepr}
\input psfig.sty

\newcommand{\ls}{\raisebox{-0.5ex}{$\,\stackrel{<}{\scriptstyle
\sim}\,$}}
\newcommand{\gs}{\raisebox{-0.5ex}{$\,\stackrel{>}{\scriptstyle
\sim}\,$}}

\begin{document}

\markboth{\sc Refregier}{\sc Weak Gravitational Lensing}
\title{Weak Gravitational Lensing by \\ Large-Scale Structure\\
 \rule{\textwidth}{1mm}}

\author{{\Large Alexandre Refregier}
\affiliation{Service d'Astrophysique, CEA/Saclay, 91191 Gif sur Yvette, 
  France;\\ email: refregier@cea.fr}}

\begin{keywords}
Cosmology, Dark Matter, Cosmic Shear, Structure Formation,
Statistics
\end{keywords}

\begin{abstract}
Weak gravitational lensing provides a unique method to map directly
the distribution of dark matter in the universe and to measure
cosmological parameters. This cosmic-shear technique is based on the
measurement of the weak distortions that lensing induces in the shape
of background galaxies as photons travel through large-scale
structures. This technique is now widely used to measure the mass
distribution of galaxy clusters and has recently been detected in
random regions of the sky. In this review, we present the theory and
observational status of cosmic shear. We describe the principles of
weak lensing and the predictions for the shear statistics in favored
cosmological models. Next, we review the current measurements of
cosmic shear and show how they constrain cosmological parameters. We
then describe the prospects offered by upcoming and future
cosmic-shear surveys as well as the technical challenges that
have to be met for the promises of cosmic shear to be fully realized.
\end{abstract}

\maketitle

\section{INTRODUCTION}
Gravitational lensing provides a unique method to directly map the
distribution of dark matter in the universe. It relies on the
measurement of the distortions that lensing induces in the images of
background galaxies (for reviews, see Bartelmann \& Schneider 1999,
Bernardeau 1999, Kaiser 1999, Mellier 1999, Narayan \& Bartelmann
1996, Schneider 1995, Wittman 2002). This method is now widely used to
map the mass of clusters of galaxies (see Fort \& Mellier 1994 for a
review) and has been extended to the study of superclusters (Gray et
al. 2002, Kaiser et al. 1998) and groups (Hoekstra et
al. 2001). Recently, weak lensing was statistically detected for the
first time in random patches of the sky (Brown et al. 2003;
Bacon et al. 2002; Bacon, Refregier \& Ellis 2000; Hamana et
al. 2003; H\"{a}mmerle et al. 2002; Hoekstra et al. 2002a; Hoekstra,
Yee \& Gladders 2002a; Jarvis et al.  2002; Kaiser, Wilson
\& Luppino 2000; Maoli et al. 2001; Refregier, Rhodes, \& Groth 2002;
Rhodes, Refregier \& Groth 2001; van Waerbeke et al. 2000; van
Waerbeke et al. 2001a; van Waerbeke et al. 2002; Wittman et
al. 2000). These cosmic shear surveys provide direct measurements of
large-scale structure in the universe and, therefore, of the
distribution of dark matter. Unlike other methods that probe the
distribution of light, weak lensing measures the mass and can thus be
directly compared to reliable theoretical models of
structure formation. Cosmic shear can therefore be used to measure
cosmological parameters in the context of these models, thereby
opening wide prospects for cosmology (Bernardeau, van Waerbeke \&
Mellier 1997; Hu \& Tegmark 1999; Jain \& Seljak 1997; Kaiser 1998).

In the present review, we describe the theoretical and observational
status of cosmic shear. Earlier reviews of this fast-evolving field
can be found in Hoekstra et al. (2002b), Mellier et
al. (2001), van Waerbeke et al. (2002), Wittman (2002). Here, we first
describe the principles of weak lensing (Section 2). We then summarize
the different statistics used to measure cosmic shear and describe how
they are used to constrain cosmological parameters (Section 3). In
Section 4, we survey the different methods used to derive the lensing
shear from the shapes of background galaxies. We present, in Section
5, the current observations and their cosmological
significance. Future cosmic-shear surveys and the prospects they offer
for cosmology are described in Section 6.  In Section 7, we outline
how systematic effects present challenges that must be met for the
potential of these future surveys to be fully realized. We summarize
our conclusions in {Section 8}.

\section{THEORY}
\label{theory}
The idea of cosmic shear can be traced back to a lecture given by
Richard Feynman at Caltech in 1964 (J.E. Gunn, personal
communication). Several theorists (e.g., Gunn 1967,
Jaroszy\'{n}sky et al. 1990, Kristian \& Sachs 1966, Lee \&
Paczy\'{n}sky 1990, Schneider \& Weiss 1988) then studied the
propagation of light in an inhomogeneous
universe. Predictions for the statistics of the weak-lensing
distortions were then computed in a modern cosmological context by
several groups (Babul \& Lee 1991, Blandford et al. 1991, Kaiser 1992,
Miralda-Escud\'{e} 1991, Villumsen 1996). More recently, the power of
cosmic shear to measure cosmological parameters was the object of many
theoretical studies (Bernardeau, van Waerbeke \& Mellier 1997; Jain \&
Seljak 1997; Hu \& Tegmark 1999; Kaiser 1998; Kamionkowski et
al. 1997; van Waerbeke, Bernardeau \& Mellier 1999). In this section,
we briefly describe the principles of weak lensing and show how this
technique can be used to map the dark matter in the universe.

\begin{figure}
\centerline{\psfig{figure=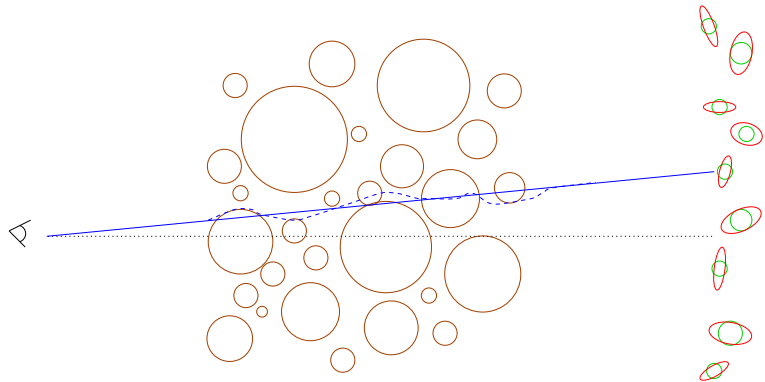,height=50mm}}
\caption{Illustration of the effect of weak lensing by large-scale
structure. The photon trajectories from distant galaxies ({\it right})
to the observer ({\it left}) are deflected by intervening large-scale
structure ({\it center}). This results in coherent distortions in the
observed shapes of the galaxies. These distortions, or shears, are
on the order of a few percent in amplitude and can be measured to
yield a direct map of the distribution of mass in the universe.}
\label{fig:lensing}
\end{figure}

\begin{figure}
\centerline{\psfig{figure=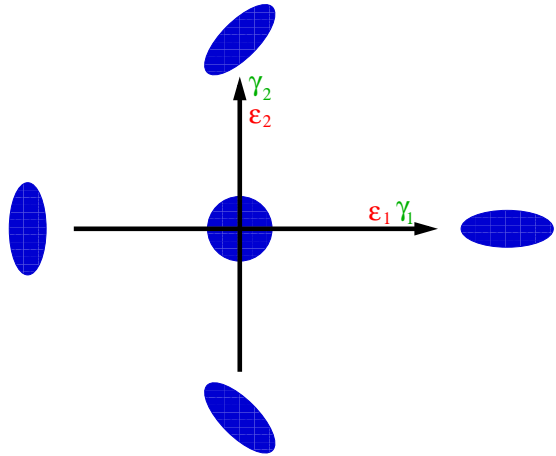,height=60mm}}
\caption{Illustration of the geometrical meaning of the shear
$\gamma_{i}$ and of the ellipticity $\epsilon_{i}$.  A positive
(negative) shear component $\gamma_{1}$ corresponds to an elongation
(compression) along the $x$-axis. A positive (negative) value of the
shear component $\gamma_{2}$ corresponds to an elongation
(compression) along the $x=y$ axis. The ellipticity of an object is
defined to vanish if the object is circular ({\it center}). The
ellipticity components $\epsilon_{1}$ and $\epsilon_{2}$ correspond
to compression and elongations similar to those for the shear components.}
\label{fig:ellip}
\end{figure}

As they travel from a background galaxy to the observer, photons
get deflected by mass fluctuations along the line of sight (see
Figure~\ref{fig:lensing}). As a result, the apparent images of
background galaxies are subject to a distortion that is
characterized by the distortion matrix:
\begin{equation}
\label{eq:psi_def_theory} \Psi_{ij} \equiv \frac{\partial (\delta
\theta_{i})}{\partial \theta_{j}}
  \equiv \left( \begin{array}{cc}
\kappa +\gamma_{1} & \gamma_{2}\\ \gamma_{2} & \kappa - \gamma_{1}
\\
\end{array} \right),
\end{equation}
where $\delta \theta_{i}({\mathbf \theta})$ is the deflection vector
produced by lensing on the sky. The convergence $\kappa$ is
proportional to the projected mass along the line of sight and
describes overall dilations and contractions. The shear $\gamma_{1}$
($\gamma_{2}$) describes stretches and compressions along (at
$45^{\circ}$ from) the x-axis. Figure~\ref{fig:ellip} illustrates the
geometrical meaning of the two shear components.

The distortion matrix is directly related to the matter density
fluctuations along the line of sight by
\begin{equation}
\label{eq:psi_dchi}
\Psi_{ij} = \int_{0}^{\chi_{h}} d\chi ~g(\chi)
\partial_{i}
   \partial_{j} \Phi,
\end{equation}
where $\Phi$ is the Newtonian potential, $\chi$ is the comoving
distance, $\chi_{h}$ is the comoving distance to the horizon, and
$\partial_{i}$ is the comoving derivative perpendicular to the
line of sight. The radial weight function $g(\chi)$ is given by
\begin{equation}
g(\chi) = 2 \int_{\chi}^{\chi_{h}} d\chi'~n(\chi')
   \frac{r(\chi)r(\chi'-\chi)}{r(\chi')},
\end{equation}
where $r=a^{-1}D_{A}$, and $D_{A}$ is the angular-diameter
distance. The function $n(\chi)$ is the distribution of the galaxies
as a function of the comoving distance $\chi$ from the observer and is
assumed to be normalized as $\int d\chi n(\chi)=1$.

\begin{figure}
\centerline{\psfig{figure=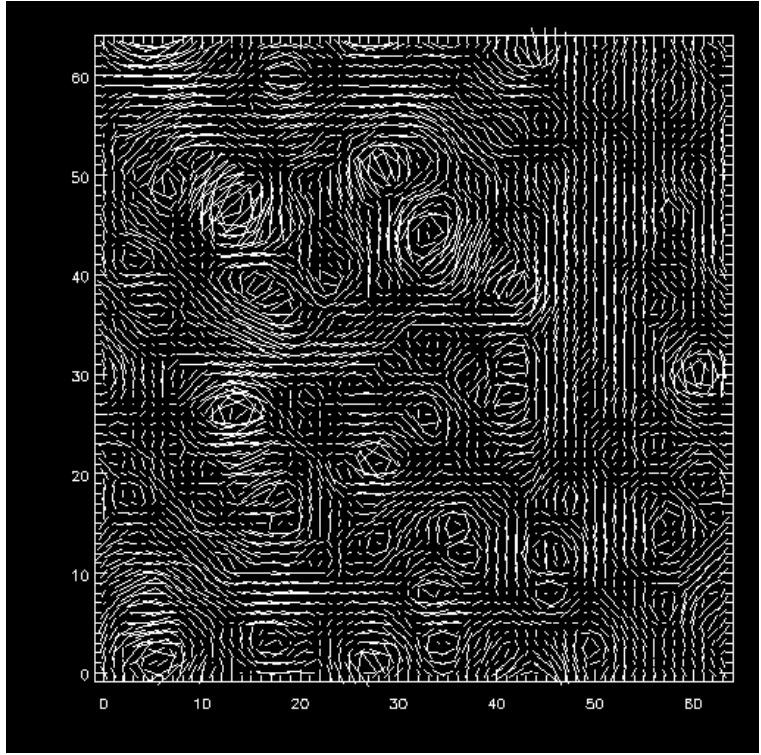,height=100mm}}
\caption{Shear map derived by ray-tracing simulations by Jain, Seljak
\& White (2000). The size and direction of each line gives the
amplitude and position angle of the shear at this location on the sky.
The displayed region is $1^{\circ}\times 1^{\circ}$ for an SCDM
(Einstein-De Sitter) model.  Tangential patterns about the
overdensities corresponding to clusters and groups of galaxies are
apparent. A more complex network of patterns is also visible outside
of these structures. The root-mean-square shear is approximately 2\%
in this map. (From Jain et al. 2000)}
\label{fig:jsw}
\end{figure}

As we discuss in Section 4, galaxy shapes can be averaged over a patch
of the sky to measure the shear, which is thus an observable. The
shear pattern expected in a standard Cold Dark Matter (SCDM)
model is shown in Figure~\ref{fig:jsw} for a $1\times1$ deg$^{2}$
region. Jain, Seljak, \& White (2000) derived this map from
ray tracing through N-body simulations. Tangential patterns around the
overdensities corresponding to clusters and groups of galaxies, along
with a more complicated network of shear fluctuations, are
apparent. By inverting the lensing equation
(Equation~\ref{eq:psi_dchi}), the shear map can be converted into a
map of the projected mass $\kappa$ and, therefore, of the dark matter
distribution.

\section{COSMIC-SHEAR STATISTICS AND COSMOLOGY}
\label{statistics}
\begin{figure}
\centerline{\psfig{figure=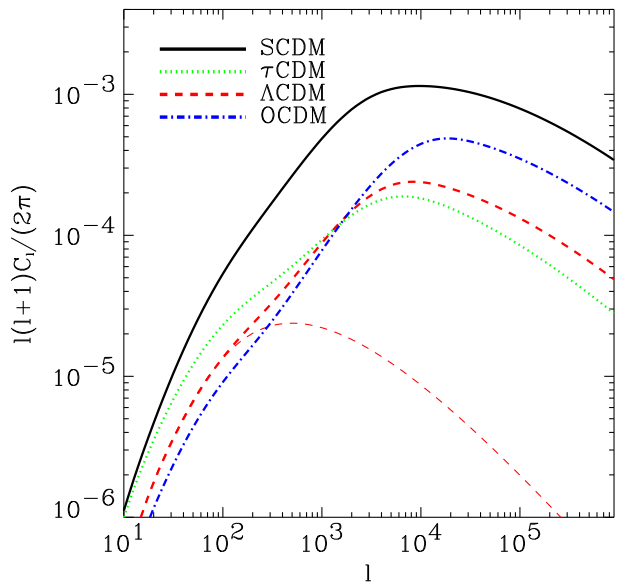,width=90mm}}
\caption{Shear power spectrum for different cosmological models and
for source galaxies at $z_{s}=1$. The SCDM model is COBE
normalized and thus has a higher amplitude than the three
cluster-normalized models $\Lambda$CDM, OCDM, and $\tau$CDM.  The thin
dashed line shows the $\Lambda$CDM spectrum for linear evolution of
structures. Notice that for $l > 1000$ (corresponding approximately to
angular scales $\theta < 10'$) the lensing power spectrum is dominated
by nonlinear structures.}
\label{fig:cl}
\end{figure}

The statistical characteristics of the cosmic-shear field can be
quantified using a variety of measures, which can then be used to
constrain cosmological models. First, we consider the most basic
two-point statistic of the shear field, namely the
two-dimensional power spectrum (Jain \& Seljak 1997, Kaiser 1998,
Kamionkowski et al. 1997, Schneider et al. 1998a). The shear power
spectrum $C_{{\mathbf l}}$ is defined as a function of
multipole moment $l$ (or inverse angular scale) by $\sum_{i=1}^{2}
\langle \widetilde{\gamma_{i}}({\mathbf l})
\widetilde{\gamma_{i}}({\mathbf l'}) \rangle = (2\pi)^{2}
\delta({\mathbf l} - {\mathbf l'}) C_{{\mathbf l}}$, where tildes
denote Fourier transforms (with the conventions of Bacon, Refregier,
\& Ellis 2000), $\delta$ is the two-dimensional Dirac-delta function,
and the brackets denote an ensemble average.  Applying Limber's
equation in Fourier space (e.g., Kaiser 1998) to
Equation~\ref{eq:psi_dchi} and using the Poisson equation, one can
easily express the shear power spectrum $C_{l}$ in terms of the
three-dimensional power spectrum $P(k,\chi)$ of the mass fluctuations
$\delta \rho/\rho$ and obtain
\begin{equation}
C_{l} = \frac{9}{16} \left( \frac{H_{0}}{c} \right)^{4}
\Omega_{m}^{2}
  \int_{0}^{\chi_h} d\chi~\left[ \frac{g(\chi)}{a r(\chi)} \right]^{2}
  P\left(\frac{l}{r}, \chi\right),
\end{equation}
where $a$ is the expansion parameter, and $H_{0}$ and $\Omega_{m}$ are
the present value of the Hubble constant and matter density parameter,
respectively. The lensing power spectra for four CDM models are shown
in Figure~\ref{fig:cl} (see color insert) (see Bacon, Refregier \&
Ellis 2000 for the exact cosmological parameter values of each
model). They were derived using the fitting formula for the nonlinear
matter power spectrum $P(k,\chi)$ of Peacock \& Dodds (1996). In
Figure 4, the galaxies were assumed to lie on a sheet at a redshift
$z_{s}=1$. A more realistic redshift distribution $n(z)$ would require
corrections of only approximately 10\% on these power spectra, as long
as the median redshift of the galaxies were kept at $z_{s}=1$.

The three cluster-normalized models ($\Lambda$CDM, OCDM and
$\tau$CDM) yield power spectra of similar amplitudes but with
different shapes. The COBE-normalized SCDM model has more power on
a small scale and thus yields a larger normalization. For the
$\Lambda$CDM model, the power spectrum corresponding to a linear
evolution of structures is also shown in Figure 4 for comparison.
For $l \gs 1000$ (corresponding to angular scales smaller than
approximately 10'), nonlinear corrections dominate the power
spectrum (Jain \& Seljak 1997), making cosmic-shear sensitive to
gravitational instability processes.

The measurement of the lensing power spectrum can thus be used to
measure cosmological parameters, such as $\Omega_{m}$,
$\Omega_{\Lambda}$, $\sigma_{8}$, and $\Gamma$. (Bernardeau, van
Waerbeke \& Mellier 1997; Hu \& Tegmark 1999; Jain \& Seljak 1997;
Kaiser 1998; van Waerbeke, Bernardeau \& Mellier 1999). A full-sky
cosmic-shear survey would yield a precision of these parameters
comparable to that for future cosmic microwave background (CMB)
missions. More realistically for the short term, a precision on the
order of 10\% can be achieved with surveys of approximately 10 square
degrees. Such cosmic-shear surveys can also be combined with CMB
anisotropy measurements to break degeneracies present when the CMB
is considered alone (Hu \& Tegmark 1999; Contaldi, Hoekstra
\& Lewis 2003). This would yield improvements in the precision of
cosmological parameters by approximately one order of magnitude. In
addition, the use of photometric redshifts can provide a tomographic
measurement of matter fluctuations and improve the precision of
cosmological parameters by up to an order of magnitude (Hu 1999, Hu \&
Keeton 2002, Taylor 2001).

In practice, it often is more convenient to measure other two-point
statistics. In particular, the variance of the shear
$\sigma_{\gamma}^{2} \equiv \langle \overline{\gamma}^{2} \rangle$
in randomly placed cells is widely used. It is related to the shear
power spectrum by
\begin{equation}
\sigma_{\gamma}^{2} = \frac{1}{2\pi} \int_{0}^{\infty}
  dl~ l C_{l} \left| \widetilde{W}_{l} \right|^{2},
\label{eq:sigma_gamma}
\end{equation}
where $\widetilde{W}_{l}$ is the Fourier transform of the cell
aperture. The shear two-point correlation functions (eg. Kaiser 1998)
are Fourier transforms of the power spectra and have also been
measured by various groups (eg. Bacon et al. 2002; van Waerbeke et
al. 2001a; Hoekstra et al. 2002a). The $M_{ap}$ statistics (Schneider
et al. 1998a) is another convenient statistic based on the average of
the shear within a compensated filter. Its advantages are that its
window function in $l$-space is narrow, that adjacent cells are
effectively uncorrelated, and that it can be easily related to the
statistics of the projected mass $\kappa$.

In analogy with electromagnetism, a tensor field like the shear field
$\gamma_{i}$ can be decomposed into an electric (E, or gradient)
component and a magnetic (B, or curl) component (Crittenden et
al. 2001b, Kaiser 1992, Kamionkowski et al. 1997, Stebbins
1996). Because the distortions it induces arise from a scalar field
(the gravitational potential), weak lensing only produces E-type
fluctuations.  On the other hand, systematic effects and intrinsic
galaxy alignments are likely to produce both E-type and B-type
fluctuations (See Heavens 2001 for a review; also see Section
7.6). The presence of B-modes can thus be used as a measure of
contaminants to the cosmic-shear signal. In practice, this
decomposition can be performed using the $M_{ap}$ statistic (Schneider
et al. 1998a, van Waerbeke et al. 2001a) or the power spectrum (Hu \&
White 2000, Padmanabhan, Seljak \& Pen 2002, Pen et
al. 2001).
\label{eb}

As is apparent in Figure~\ref{fig:jsw}, the shear field is not
Gaussian but, instead, has apparent coherent structures. Bernardeau,
van Waerbeke, \& Mellier (1997) have shown that the skewness
\begin{equation}
S_{3} = \frac{\langle \kappa^{3} \rangle}{\langle \kappa^{2} \rangle^2}
\end{equation}
of the convergence field $\kappa$ can be used to break the degeneracy
between $\sigma_{8}$ and $\Omega_{m}$, which is present when the shear
variance alone is considered. The measurement of higher-point
statistics is thus of great cosmological interest, but their
computation is made difficult by the fact that the angular scales
accessible to obervation ($\theta \ls 10'$) are in the
nonlinear regime. Predictions for higher-point correlation functions
have been calculated using perturbation theory, the hierarchical
ansatz, (Bernardeau \& Valageas 2000; Bernardeau, van Waerbeke \&
Mellier 1997; Hui 1999; Munshi \& Coles 2000; Munshi \& Jain 2001; van
Waerbeke et al. 2001b) and halo models (Cooray \& Hu 2001a; Cooray,
Hu, \& Miralda-Escud\'{e} 2000). These techniques have also been used
to compute the full probability distribution function of the
convergence field (Bernardeau \& Valageas 2000, Munshi \& Jain 2000,
Valageas 2000) and the errors in two-point statistics (Cooray \& Hu
2001b; Munshi \& Coles 2002; Schneider et al. 2002).

Another possible way to compute the full nonlinear field is to use ray
tracing through N-body simulations (Blandford et al. 1991;
Hamana, Colombi \& Mellier 2000; Jain, Seljak \& White 2000; Premadi
et al. 2001; Wambsganss, Cen, \& Ostriker 1998; White \& Hu 2000) to
produce simulated shear maps such as those by Jain, Seljak \& White
(2000) (shown in Figure~\ref{fig:jsw}). These can be used to compute
and study other proposed measures of non-Gaussianity such as peak
statistics (Jain \& van Waerbeke 2000), a generalized maximum
likelihood (Taylor \& Watts 2000), and cluster counts (Bartelmann,
King \& Schneider 2001). In practice, the complex geometry of surveys
makes it difficult to infer a convergence $\kappa$ map from the
observed shear $\gamma_{i}$. For this reason, a number of researchers
have recently proposed the use of high-order statistics of the shear
field rather than the skewness $S_{3}$ of the convergence (Bernardeau,
van Waerbeke, \& Mellier 2003; Schneider \& Lombardi 2002; Takada \&
Jain 2002; Zaldarriaga \& Scoccimarro 2002).

\section{SHEAR MEASUREMENT METHODS}
\label{method} Because the sought-after lensing signal is of only
a few percent in amplitude, the data acquisition and analysis must be
performed carefully, and systematic effects must be tightly
controlled. All the current measurements of cosmic shear were derived
from deep optical images taken with charged-coupled devices
(CCD). It is advantageous for the exposures to be homogeneous in
depth and for the ground to be subject to as small a seeing as
possible. The fields are generally chosen to lie far away from each
other to ensure that they are statistically independent and to
minimize cosmic variance.

\begin{figure}
\centerline{\psfig{figure=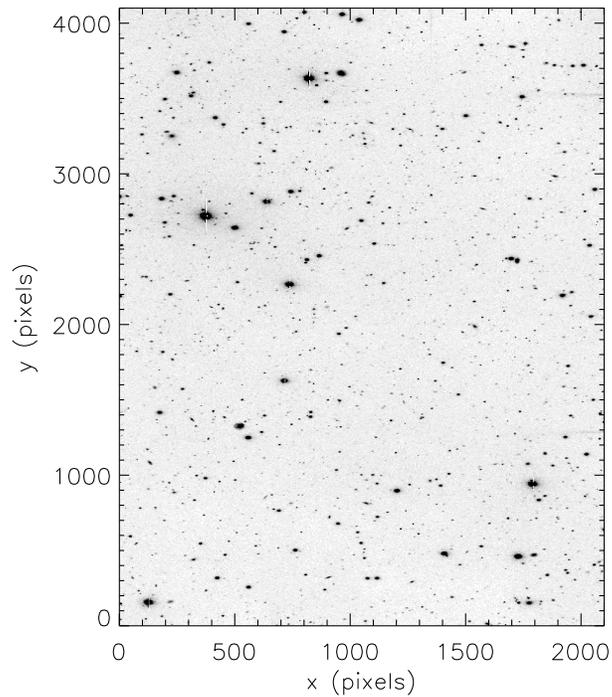,height=100mm}} 
\caption{Example of an deep image in the cosmic-shear survey by Bacon,
Refregier \& Ellis (2000). This corresponds to a 1 h exposure
with the EEV camera on the William Herschel Telescope (WHT). The field
of view is $8'\times 16'$ and achieves a magnitude depth of
$R\simeq26$ (5$\sigma$ detection). The bright objects are saturated
stars. The faint objects comprise approximately 200 stars and
approximately 2000 galaxies that are usable for the weak-lensing
analysis. (From Bacon, Refregier \& Ellis 2000)}
\label{fig:wht}
\end{figure}

The first step in the data analysis is image processing. After flat
fielding, the different exposures are co-added to produce the final
reduced images. If necessary, any instrumental distortion
induced by the telescope optics is corrected for at this
stage.  This can be done very accurately by measuring the astrometric
offsets from several dithered exposures. For a detailed
description of the different image-processing steps, see
Kaiser et al. (1999). An example of a processed deep image from the
cosmic-shear survey of Bacon, Refregier \& Ellis (2000) is shown in
Figure~\ref{fig:wht}.

\begin{figure}
\centerline{\psfig{figure=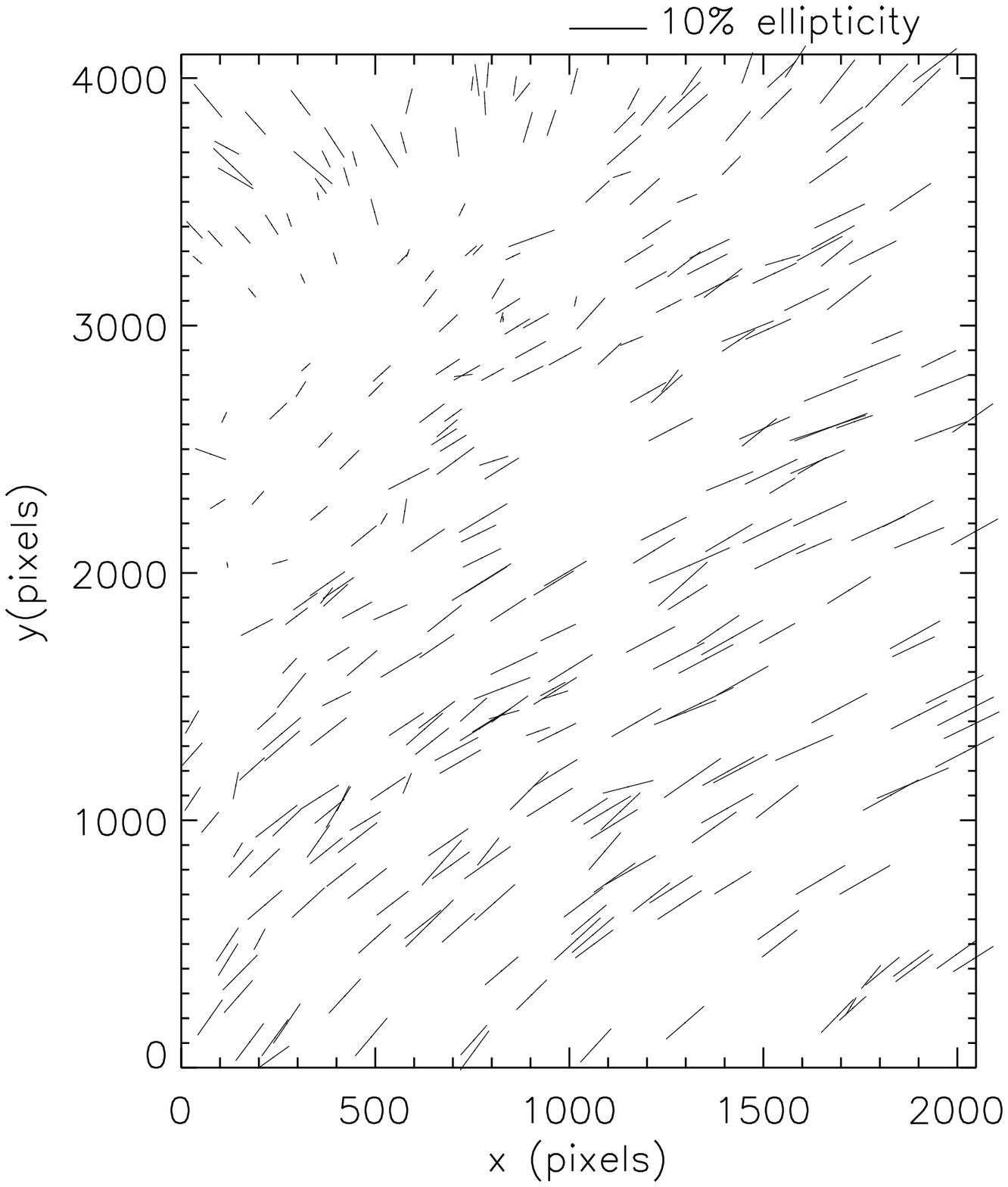,height=80mm}
  \psfig{figure=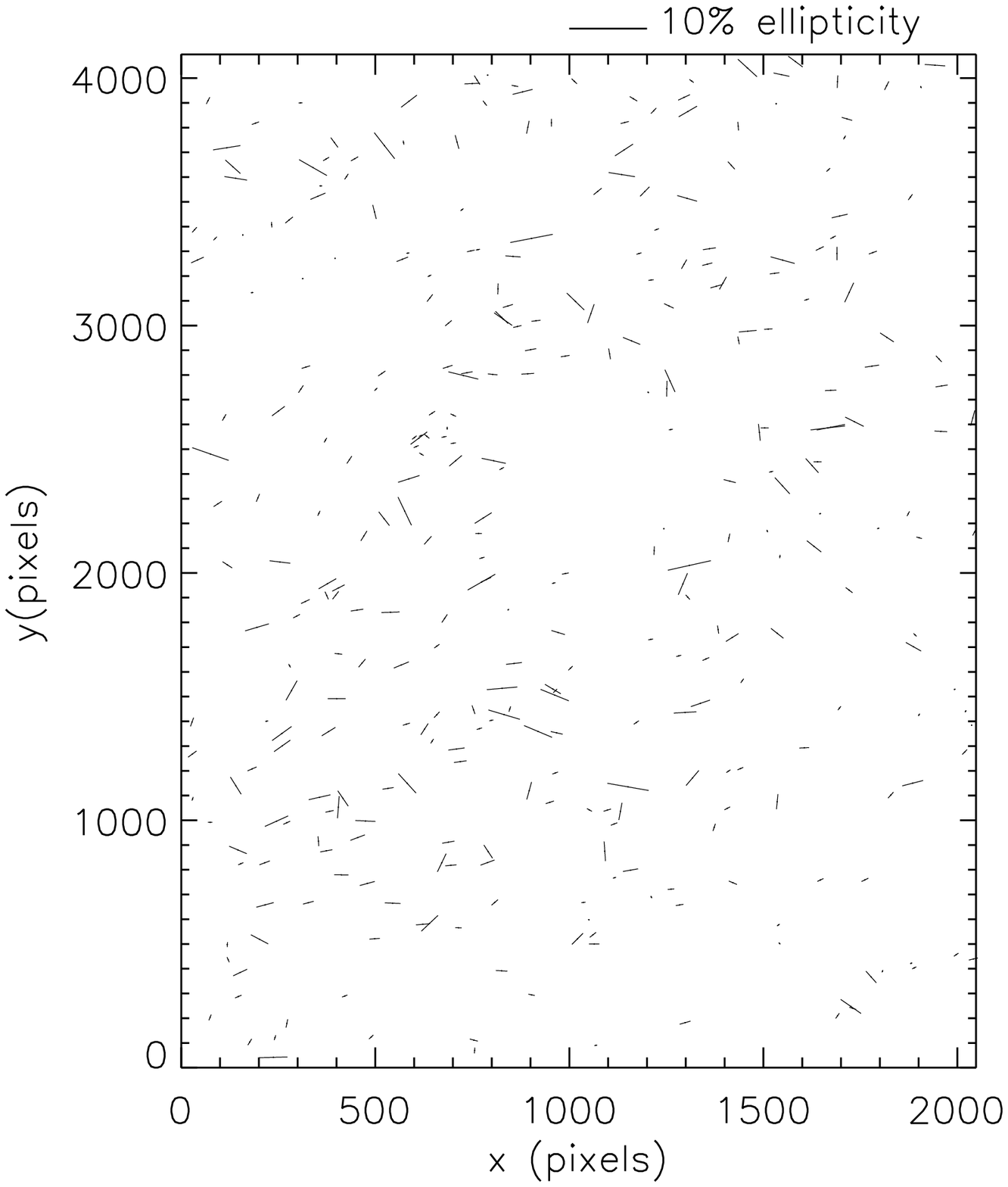,height=80mm}}
\caption{PSF ellipticity pattern measured in WHT field shown in the
Figure 5: ({\it a}) PSF ellipticities measured from the stars in
Figure 5. ({\it b}) Ellipticity residuals for those stars after the
KSB corrections.  In this survey, the root-mean-square stellar
ellipticity is found to be approximately 7\% before the corrections
and negligible after the correction. (From Bacon, Refregier \& Ellis
2000)}.
\label{fig:star}
\end{figure}

The next step consists of deriving an estimator for the shear from the
shapes of the galaxies in the co-added images. The point-spread
function (PSF), which smears the images of galaxies and is
generally not circular, complicates this task.  In general, the PSF
varies spatially and in time, and it must be measured and
modelled for each image individually. This can be done by
measuring the shape of the stars in the field, whose number can be
optimized by tuning the galactic latitudes of the observations.
Figure~\ref{fig:star} shows the ellipticity pattern of the PSF from
one of the William Hershel telescope (WHT) fields of Bacon, Refregier
\& Ellis 2000.

Several methods have been developed to tackle this difficult and
crucial task. The more rigorous method of Kaiser, Squires \&
Broadhurst (KSB; 1995), further developed by Luppino \& Kaiser (1997)
and Hoekstra et al. (1998), replaced the earlier method by
Bonnet \& Mellier (1995). The KSB method is now widely used
for cluster studies, and it has been used by the majority of the
groups involved in measuring cosmic shear (see Bartelmann \& Schneider
1999 for a detailed review of the KSB method). It is based on the
measurement of the quadrupole moment of the galaxy surface brightness
$I({\mathbf x})$,
\begin{equation}
Q_{ij} \equiv \int d^{2}x x_{i} x_{j} w(x) I({\mathbf x}),
\end{equation}
where $w(x)$ is a weight function conveniently taken to be
Gaussian. These moments capture the lowest-order shape information of
the galaxy and can be combined to form the ellipticity of the galaxy
\begin{equation}
\epsilon_{1}= \frac{Q_{11}-Q_{22}}{Q_{11}+Q_{22}},~~~
\epsilon_{2}= \frac{2Q_{12}}{Q_{11}+Q_{22}}.
\end{equation}
The first (second) component of the ellipticity describes compressions
and elongations along (at 45$^{\circ}$ from) the x and y axes (see
Figure~\ref{fig:ellip} for an illustration). The ellipticity vanishes
for circular galaxies.

The first step in the KSB method is to correct the observed galaxy
ellipticity $\epsilon_{i}^{g \prime}$ for the anisotropy of the
PSF. The corrected galaxy ellipticity $\epsilon^{g}$ is given by
\begin{equation}
\epsilon^{g} = \epsilon^{g \prime} - P_{\rm sm}^{g} (P_{\rm
sm}^{*})^{-1} \epsilon^{*},
\end{equation}
where $\epsilon^{*}$ is the PSF ellipticity derived from the stars,
and $P_{\rm sm}^{g}$ and $P_{\rm sm}^{*}$ are the smear susceptibility
tensors for the galaxy and star, respectively, and can be derived from
higher moments of the images. The shear in a patch of the sky can then
be measured by averaging over the (corrected) ellipticities in the
patch of the sky using
\begin{equation}
\gamma = (P_{\gamma})^{-1} \langle \epsilon^{g} \rangle,
\end{equation}
where the tensor $P_{\gamma}$ quantifies the susceptibility to shear
acting before isotropic PSF smearing and is given by
\begin{equation}
P_{\gamma} = P_{\rm sh}^{g} - P_{\rm sh}^{*} (P_{\rm sm}^{*})^{-1}
  P_{\rm sm}^{g},
\end{equation}
where the shear susceptibility tensors $P_{\rm sh}^{g}$ and
$P_{\rm sh}^{*}$ for the galaxies and the stars can also be
measured from higher moments of their respective light
distribution.

The KSB method was thoroughly tested using realistic simulated images
by Erben et al. (2001) and Bacon et al. (2001a). These studies showed
that it is accurate to within a few tenths of percent in
reconstructing an input shear and is thus sufficient for the current
cosmic-shear surveys (see Section 5). However, this accuracy is
insufficient for future, more sensitive surveys (see Section
6). Moreover, Kuijken (1999) and Kaiser (2000) have shown that the KSB
method is ill-defined mathematically and unstable for PSF's found in
practice.

This inadequacy has led a number of researchers to develop alternative
methods.  Rhodes, Refregier \& Groth (2000) have modified the KSB
method to be better suited for HST images. Kuijken (1999)
considered a different approach that consisted of fitting the observed
galaxy shape with a smeared and sheared circular model. Kaiser (2000)
introduced a new method based on a finite resolution shear
operator. Refregier \& Bacon (2003) and, independently, Bernstein \&
Jarvis (2001) developed a new method based on the decomposition of the
galaxies into shape components or ``shapelets''. The
gauss-hermite orthogonal basis functions used in this approach allow
shears and PSF convolutions to be described as simple matrix
operations, using the formalism developped for the quantum harmonic
oscillator (Refregier 2003). These new methods are promising but
require extensive testing to establish whether they will achieve the
required precision.

\section{OBSERVATIONS}
\label{observations} Because the expected distortions are only of
a few percent, the measurement of cosmic shear requires large survey
areas and excellent image quality. Early searches for cosmic
shear signals with photographic plates were unsuccesful (Kristian
1967; Valdes, Jarvis \& Tyson 1983). Mould et al. (1994), performed
the first attempt to detect a cosmic-shear signal with CCDs, but only
derived an upper limit. Using the same data, Villumsen (1995)
reported a $4.5\sigma$ detection. Schneider et al. (1998b) then
reported a detection of cosmic shear in one of three QSO fields, an
area too small to draw any constraints on cosmology.

Within a few weeks, four independent groups (Bacon, Refregier \& Ellis
2000; Kaiser, Wilson \& Luppino 2000; van Waerbeke et al.  2000;
Wittman et al. 2000) reported the first firm statistical detections of
cosmic shear using three different 4m-class telescopes: the Cerro
Tololo Inter-American Observatory (CTIO), the Canada-France-Hawaii
Telescope (CFHT), and the William Herschel Telescope (WHT). These were
later confirmed by more precise measurements of the cosmic-shear
amplitude from the ground (Bacon et al. 2002; Brown et al. 2003; Maoli
et al. 2001; Hamana et al. 2003; Hoekstra et al. 2002a; Hoekstra, Yee
\& Gladders 2002a; Jarvis et al. 2002; van Waerbeke et al. 2001; van
Waerbeke et al. 2002a) using 2-m -- (MPG/ESO), 4-m-- (WHT, CFHT, CTIO)
and 8-m-- (Very Large Telescope, Keck, Subaru) class ground-based
telescopes.  Meanwhile, cosmic shear was also detected (H\"{a}mmerle
et al. 2002; Rhodes, Refregier \& Groth 2001) and measured (Refregier,
Rhodes \& Groth 2002) from space using HST. Space-based surveys are
currently limited by the small field of view of HST, but they are
deeper and less prone to systematics thanks to the absence of
atmospheric seeing.

\begin{figure}
\centerline{\psfig{figure=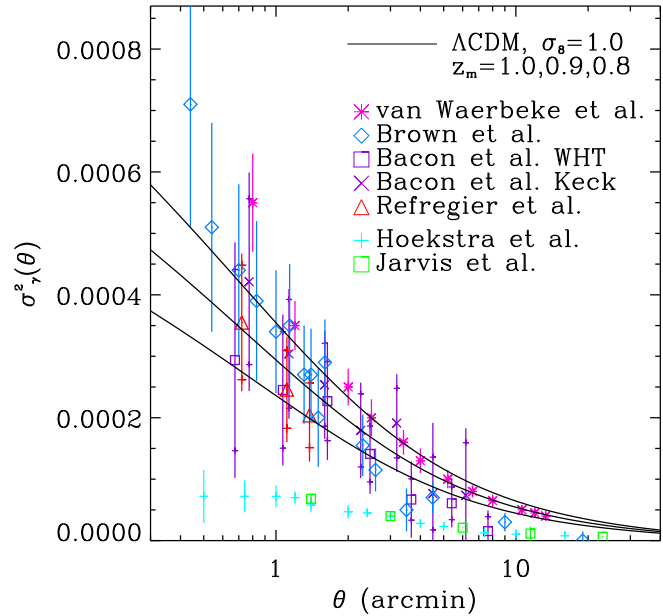,height=90mm}} 
\caption{Shear variance $\sigma_{\gamma}^2$ as a function of the
radius $\theta$ of a circular cell. The data points correspond to
recent results from different groups: van Waerbeke et al. (2002a),
Brown et al. (2003), Bacon et al. (2002, WHT and Keck), Refregier et
al.  (2002), Hoekstra et al. (2002b), Jarvis et al. (2002).  When
relevant, the inner error bars correspond to noise only, whereas the
outer error bars correspond to the total error (noise + cosmic
variance). The measurements by H\"{a}mmerle et al. (2002) and Hamana
et al. (2003) are not displayed. The solid curves show the
predictions for a $\Lambda$CDM model with $\Omega_{m}=0.3$,
$\sigma_{8}=1$, and $\Gamma=0.21$. The galaxy median redshift was
taken to be $z_{m}=1.0,0.9,$ and $0.8$, from top to bottom,
respectively, corresponding approximately to the range of depth of the
top five surveys. The bottom two surveys (Hoekstra et al. 2002b;
Jarvis et al. 2002) have a median redshift in the range $z_{m} \simeq
0.6$--$0.7$ and, as expected, yield lower shear variances.  }
\label{fig:var}
\end{figure}

Table~\ref{tab:surveys} summarizes the existing cosmic-shear surveys
and highlights the wide range of telescopes, survey areas, and
depths. The shear variance $\sigma_{\gamma}^{2}$
(Equation~\ref{eq:sigma_gamma}) measured recently by several groups is
shown in Figure~\ref{fig:var} (see color insert) as a function of the
radius a circular cell. The results by H\"{a}mmerle et al. (2002) and
Hamana et al. (2002a) are not displayed. Note that, for the shear
variance, the data points at different angular scales are not
independent. For comparison, the shear variance predicted for a
$\Lambda$CDM model with $\sigma_{8}=1$ is also shown in
Figure~\ref{fig:var}. The model is displayed for median galaxy
redshifts of $z_{m}$ from 0.8 to 1.0, corresponding approximately to
the range of depths of the top five surveys displayed (van Waerbeke et
al. 2002a; Brown et al. 2003; Bacon et al. 2002, WHT and Keck;
Refregier et al. 2002). These observations are approximately
consistent with each other and with the $\Lambda$CDM model on angular
scales from 0.7 to 20 arcmins. This is compelling given that these
were performed with different telescopes (and therefore different
instrumental systematics) and independent data-analysis pipelines.

The bottom two surveys (Hoekstra et al. 2002b; Jarvis et
al. 2002) have a median redshift in the range $z_{m} \simeq
0.6$--$0.7$ and yield lower shear variances. As indicated by the
theoretical curves in figure~\ref{fig:var}, this redshift dependence
is expected in CDM models and thus confirms the detection of a cosmic
shear signal.

\begin{table*}
 \centering
 \begin{minipage}{140mm}
 \caption{current cosmic-shear surveys}
 \label{tab:surveys}
 \begin{tabular}{lllll}
\hline
\hline
Reference & Telescope & Area (deg$^{2}$) & Mag. limit &
$\sigma_{8}$\footnote{for $\Lambda$CDM model with $\Omega_{m}=0.3$,
$\Omega_{\Lambda}=0.7$; $\Gamma$ is marginalised over or set to
$0.21$ when possible; errors correspond to 68\%CL} \\
\hline
Wittman et al. 2000 & CTIO & 1.0 & $R \ls 26$ \\
van Waerbeke et al. 2000 & CFHT & 1.7 & & \\
Kaiser et al. 2000 & CFHT & 0.96 & $I\ls 24$, $V \ls 25$ & \\
Bacon et al. 2000 & WHT & 0.5 & $R \ls 26$ & $1.50^{+0.50}_{-0.50}$ \\
Maoli et al. 2001 & VLT & 0.65 & $I \ls 24.5$ &
  $1.03^{+0.03}_{-0.03}$\footnote{for combination of Maoli et al. (2001),
  van Waerbeke et al. (2000), Bacon et al. (2000), and Wittman et al. (2000);
  cosmic variance not included.} \\
Rhodes et al. 2001 & HST/WFPC2 & 0.05 & $I \ls 26$ & $0.91^{+0.25}_{-0.30}$ \\
van Waerbeke et al. 2001a & CFHT & 6.5 & $I \ls 24.5$ &
  $0.88^{+0.02}_{-0.02}$\footnote{cosmic variance not included.} \\
H\"{a}mmerle et al. 2002 & HST/STIS & 0.02 & & \\
Hoekstra et al. 2002a & CFHT, CTIO & 24 & $R \ls 24$ &
  $0.81^{+0.07}_{-0.09}$ \\
van Waerbeke et al. 2002a & CFHT & 8.5 & $I \ls 24.5$ &
  $0.98^{+0.06}_{-0.06}$ \\
Refregier et al. 2002 & HST/WFPC2 & 0.36 & $\langle I \rangle \simeq 23.5$ &
   $0.94^{+0.14}_{-0.14}$ \\
Bacon et al. 2002 & WHT, Keck & 1.6 & $R \ls 26$ & $0.97^{+0.13}_{-0.13}$ \\
Hoekstra et al. 2002b & CFHT, CTIO & 53 & $R \ls 24$ &
  $0.86^{+0.04}_{-0.05}$ \\
Brown et al. 2003 & MPG/ESO & 1.25 & $R \ls 25$ & 
  $0.72^{+0.09}_{-0.09}$\\
Hamana et al. 2003 & Subaru & 2.1 & $R \ls 26$ & 
  $0.69^{+0.18}_{-0.13}$\\
Jarvis et al. 2002 & CTIO & 75 & $R \ls 23$ &
  $0.71^{+0.06}_{-0.08}$\\
\hline
\end{tabular}
\end{minipage}
\end{table*}

\begin{figure}
\centerline{\psfig{figure=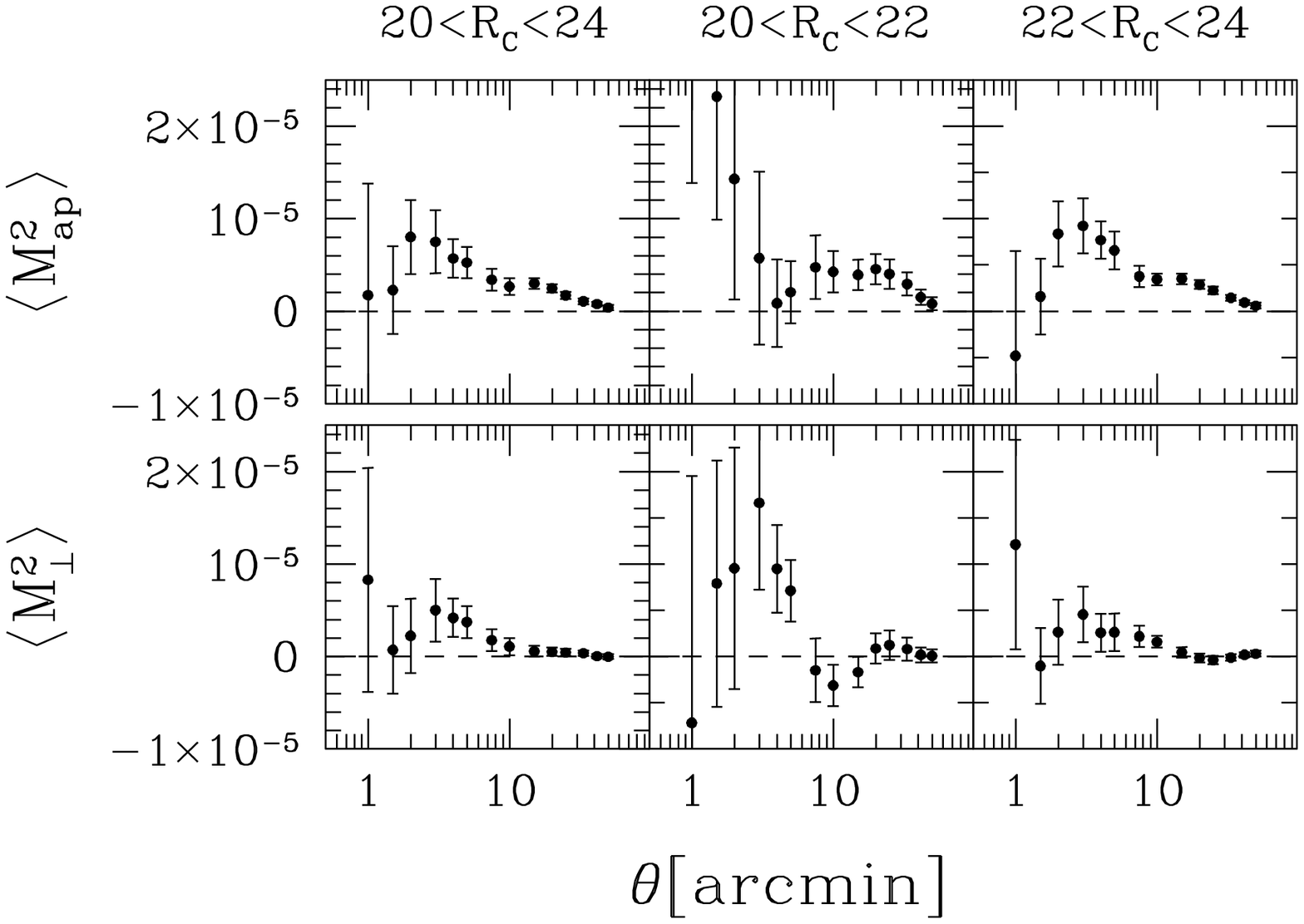,height=80mm}}
\centerline{\psfig{figure=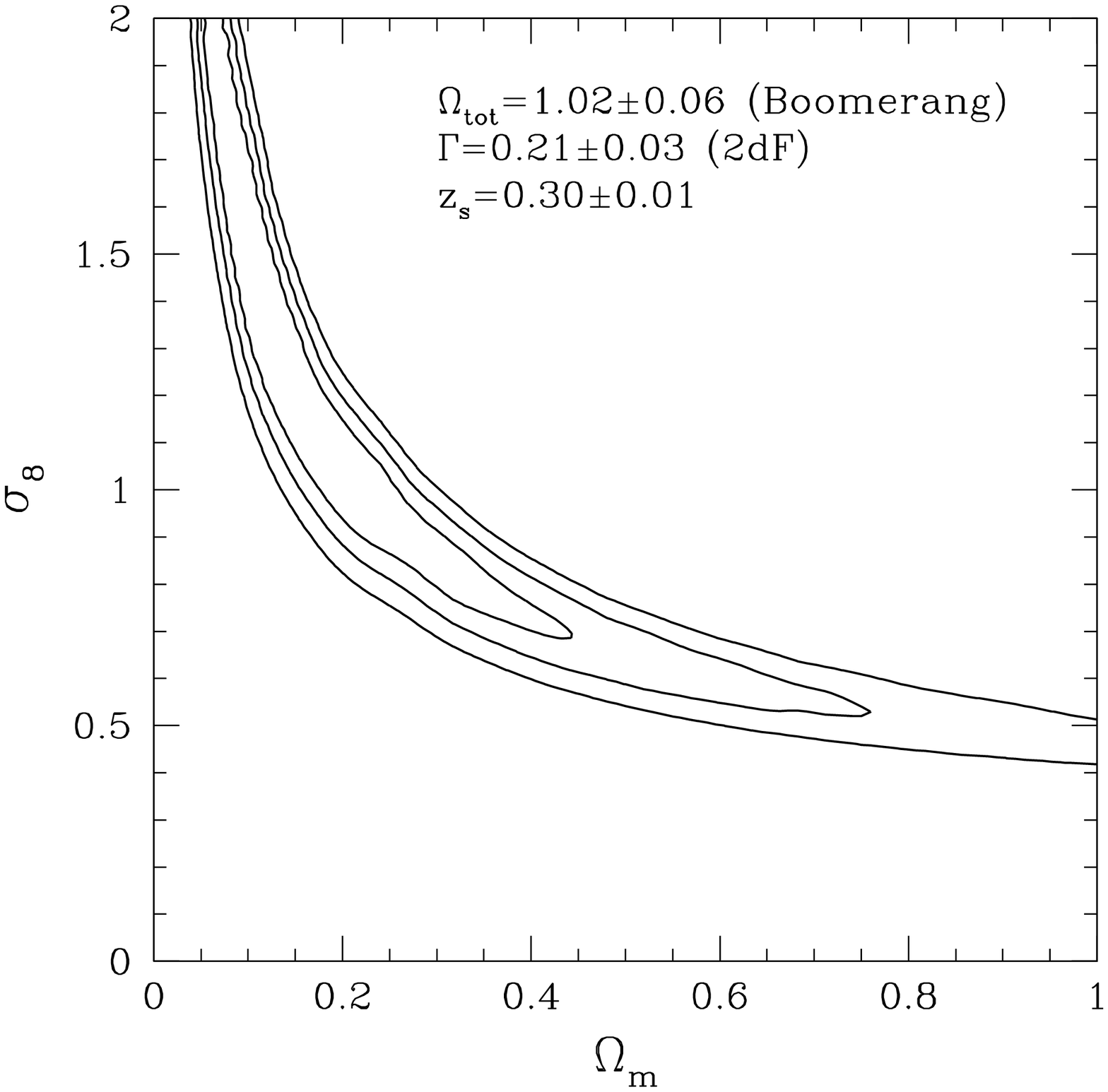,height=70mm}}
\caption{Cosmic shear measurement by Hoekstra,
Yee \& Gladders (2002a) from the RCS survey. (a): measurement of the 
$E$ (top panels) and $B$ (bottom panels) modes of the $M_{ap}$ statistics
variance as a function of aperture scale $\theta$. Several ranges
of the $R_{c}$-magnitude are considered. The errors correspond to
$1\sigma$ statistical uncertainties. The presence of significant
$B$ Modes reveal the presence of residual systematics on small
scales ($\theta \le 10'$). (b): Cosmological constraints derived
from their measurement of the $E$-mode $M_{ap}$ statistics for
$22<R_{c}<24$ after accounting for the residual $B$-modes. The
joint constraints are shown for $\Omega_{m}$ and $\sigma_{8}$ in
the $\Lambda$CDM model. Priors from CMB and galaxy surveys were
used to marginalize  over $\Omega_{\rm tot}$, $\Gamma$, and the
source redshift $z_{s}$. The contours indicate 68.3\%, 95.4\%, and
99.9\% confidence levels. (From Hoekstra, Yee \& Gladders 2002a)
\label{fig:hoek}}
\end{figure}

Although the shear variance was displayed for the purpose of
Figure~\ref{fig:var}, some of the groups have used instead the shear
correlation function or $M_{ap}$ statistic to quantify their lensing
signal. Recently, several groups used the latter statistic to separate
their signal into $E$ and $B$ components (see Section 3) and thus to
estimate and subtract systematic effects (Jarvis et al.
2002; Hamana et al. 2003; Hoekstra, Yee \& Gladders 2002; van Waerbeke
et al. 2001, 2002a). Pen, van Waerbeke \& Mellier (2001) and
Brown et al. (2003) chose instead to measure directly the shear power
spectrum for each $E$ and $B$ mode. Figure~\ref{fig:hoek}a
shows the measurement of the $E$ and $B$ signals using the $M_{ap}$
statistic by Hoekstra, Yee \& Gladders (2002a), for several
$R_{c}$-magnitude ranges in the Red-Sequence Cluster (RCS)
survey. They measure a clear lensing signal apparent as $E$
modes. However, the significant $B$ modes reveals the presence of
residual systematics on small scales ($\theta \ls 10'$), especially
for bright galaxies. In their analysis, these authors use the
amplitude of the $B$ modes as a measure systematic uncertainties.

Existing cosmic-shear measurements already yield interesting
constraints on the amplitude of the matter power spectrum $\sigma_{8}$
on which the lensing signal strongly depends.  For instance,
Figure~\ref{fig:hoek}b shows cosmological constraints for a
$\Lambda$CDM model derived by Hoekstra, Yee \& Gladders (2002a) from
the measurement of the $M_{ap}$ statistic for their $22<R_{c}<24$
galaxy sample. The degeneracy between $\sigma_{8}$ and $\Omega_{m}$
apparent in the figure is typical of cosmic-shear measurements
involving only two-point statistics. Hoekstra, Yee \& Gladders (2002a)
found that the constraints are well described by $\sigma_{8}
\Omega_{m}^{0.52}=0.46^{+0.05}_{-0.07}$ (95\% CL), where priors from
CMB and galaxy survey data have been used to marginalize over $\Gamma$
and $\Omega_{\Lambda}$. Other surveys find constraints of the same
form with similar exponents for $\Omega_{m}$.

\begin{figure}
\centerline{\psfig{figure=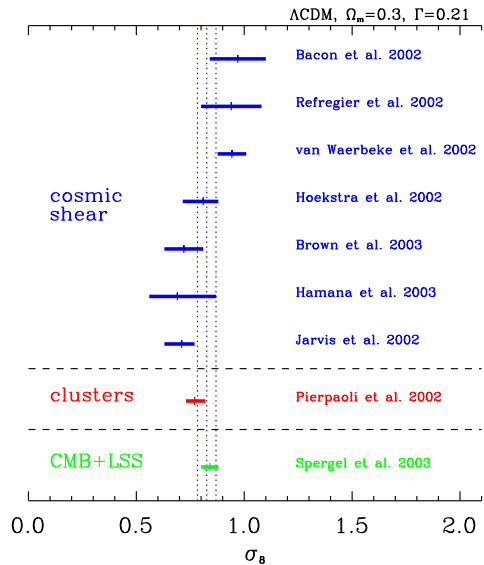,height=80mm}}
\caption{Comparison of the determination of $\sigma_{8}$ by different
groups and methods. The errors have all been converted to $1\sigma$,
and a $\Lambda$CDM model with $\Omega_{m}=0.3$ and $\Gamma=0.21$ was
assumed when possible. The vertical dotted lines show the
average $\sigma_{8} \simeq 0.83\pm0.04$ of the seven cosmic-shear
measurements and associated $1\sigma$ error. The normalization from
cluster abundance (e.g., Pierpaoli et al. 2002) as well as that
derived from CMB anisotropies and local large-scale structure (galaxy
and Lyman $\alpha$ surveys; Spergel et al. 2003) are also shown.
\label{fig:sigma8}}
\end{figure}

Table~\ref{tab:surveys} lists the values of $\sigma_{8}$ found by the
different groups. The errors have been converted to 68\% confidence
level (CL) when necessary, and values of $\Omega_{m}=0.3$,
$\Omega_{\Lambda}=0.7$, and $\Gamma=0.21$ have been assumed,
when possible. The most recent values are displayed in
Figure~\ref{fig:sigma8} (see color insert), along with their average
$\sigma_{8} \simeq 0.83 \pm 0.04$ ($1\sigma$). The values
derived by the different cosmic shear groups are in the range
$\sigma_{8}\simeq 0.7$--$1.0$, with the most recent measurements
(Brown et al. 2003; Hamana et al. 2003; Jarvis et al. 2002) yielding
lower values. The $2$--$3\sigma$ mutual inconsistencies between some
of the measurements may be symptomatic of a small level of residual
systematics. In particular, calibration errors in the shear
measurement method would not be detected via the $E$-$B$ decomposition
and are a likely explanation (see Hirata \& Seljak 2003). Another
source of discrepancy arises from the different fitting functions for
the non-linear evolution of the power spectrum (see
section~\ref{theory}). Apart from Brown et al. (2003; see also
Contaldi et al. 2003) who used the more accurate results of Smith et
al. (2003), the other groups used the Peacock \& Dodds (1997) fitting
function which yields an underprediction of $\sigma_{8}$ by roughly
5--10\%. The impact of these systematics will be discussed further in
section~\ref{challenges}.

Interestingly, an independent measurement of $\sigma_{8}$ is provided
by the abundance of X-ray clusters and can be directly compared to
this value. Initially, this method yielded normalisations of
$\sigma_{8}\sim 0.9$ (Eke et al. 1998; Pierpaoli, Scott \& White 2001;
Viana \& Liddle 1999), in agreement with the early cosmic shear
results (see table~\ref{tab:surveys} and
figure~\ref{fig:sigma8}). This value was subsequently revised downward
to $\sigma_{8}\sim 0.7$--$0.8$ by the use of the observed, rather than
simulated, mass-temperature relation in clusters (Borgani et al.
2001; Reiprich \& B\"{o}hringer 2001; Seljak 2001; Viana, Nichol \&
Liddle 2001; Pierpaoli, Scott \& White 2002). Recently, Spergel et
al. (2003) derived a value of $\sigma_{8}=0.84\pm0.04$ (68\%CL) from a
joint analysis of CMB anisotropy measurements from the Wilkinson
Microwave Anistropy Probe (WMAP) and other experiments, galaxy
clustering and the Lyman $\alpha$ forest. For comparison, these
results along with a representative value of the revised
cluster-abundance normalization (Pierpaoli, Scott \& White 2002) are
also shown in figure~\ref{fig:sigma8}.  The average of the cosmic
shear results is formally in good agreement with the determination of
$\sigma_{8}$ using the other techniques. Future surveys are however
needed to confirm this, by resolving the discrepancies between the
current cosmic shear measurements. Also, a full likelihood analysis
would be required to establish the significance of the agreement
between the different techniques (see Contaldi et al. 2003). This
comparison is important as it provides a strong test of the
$\Lambda$CDM model, the gravitational instability paradigm, the
physics of clusters, and of the biased formation of galaxies.

As discussed in Section 3, the degeneracy between $\sigma_{8}$ and
$\Omega_{m}$ can be broken by measuring higher-order correlation
functions of the lensing field. Bernardeau, Mellier, \& van
Waerbeke (2002) recently reported the first detection of a
non-Gaussian shear signal using the 3-point shear correlation function
formalism of Bernardeau, van Waerbeke \& Mellier
(2003). Another measure of non-gaussianity was recently
performed by Miyazaki et al. (2002) using peak statistics in their
Subaru survey.  Although these results are consistent with that
expected from structure formation models, larger survey areas are
needed to infer cosmological constraints.

Another approach to probe the dark matter is to measure the bias
between the mass and galaxies by cross-correlating the shear map
with that of the light from foreground galaxies in the same region
of the sky (Cooray 2002, Schneider 1998, van Waerbeke 1998).
Hoekstra, Yee \& Gladders (2001a), Hoekstra et al. (2002b), and
Wilson, Kaiser \& Luppino (2001) have recently measured this
cross-correlation on large scales.

\section{FUTURE SURVEYS AND PROSPECTS}
\label{future} The existing measurements described above are
primarily limited by statistics. They will therefore be improved upon
by ongoing surveys on existing telescopes, such as the Legacy Survey
on CFHT (CFHTLS; Mellier et al. 2000), the Deep Lens Survey (Wittman
et al.  2002), surveys with the Subaru telescope (Hiroyasu et
al. 2001) and the Sloan Digital Sky Survey (Stebbins, McKay \& Frieman
1995). Future instruments dedicated to surveys and for which cosmic
shear is a primary science driver are being planned, such as Megacam
on CFHT (Boulade et al. 2000), the Visible and Infrared Survey
Telescope for Astronomy (VISTA; Taylor et al. 2003), the Large
aperture Synoptic Survey Telescope (LSST; Tyson et al. 2002a,b), or
the novel Panoramic Survey Telescope and Rapid Response System
(Pan-STARRS; Kaiser, Tonry \& Luppino 2000). From space, the new
Advanced Camera for Surveys (ACS) on HST and, much more ambitiously,
the future Supernova Acceleration Probe satellite (SNAP; Perlmutter et
al. 2003; Rhodes et al. 2003; Massey et al. 2003; Refregier et
al. 2003) also offer exciting prospects.  Table~\ref{tab:future} lists
the characteristics of some of these future surveys. Broadly speaking,
ground-based measurements will cover large areas, whereas space-based
surveys will yield higher-resolution maps and reduced systematics
thanks to the absence of atmospheric seeing.

\begin{table*}
 \centering
 \begin{minipage}{140mm}
 \caption{Future cosmic-shear surveys}
 \label{tab:future}
 \begin{tabular}{lllllll}
\hline \hline Telescope/Survey & Ground/Space & Diameter  & FOV
& Area & Start$^{a}$ & Ref.$^{b}$ \\
& & (m) &  (deg$^{2}$) & (deg$^{2}$) & &  \\
\hline
DLS & ground & $2\times4$ & $2\times0.3$ & 28 & 1999$^{c}$ & 1 \\ 
CFHTLS  & ground & 3.6  & 1 & 172 & 2003 & 2 \\
VST & ground & 2.6 & 1 & $x$100$^{d}$ & 2004 & 3 \\
VISTA$^{e}$ & ground & 4 & 2 & 10000 & 2007 & 4 \\
Pan-STARRS & ground & $4\times1.8$ & $4\times4$ & 31000  & 2008 & 5 \\
LSST & ground & 8.4 & 7  & 30000 & 2012 & 6 \\
SNAP & space & 2 & 0.7 & 300  & 2011 & 7 \\
\hline
\end{tabular}
{\footnotesize
$^{a}$ planned start date of the cosmic shear surveys\\ $^{b}$
references: 1: Wittman et al. 2002; 2: Mellier et al. 2000; 
3: Kuijken et al. 2002; 4: Taylor et al. 2003; 5:
Kaiser et al. 2000; 6: Tyson et al. 2002a,b; 7: Perlmutter et
al. 2003, Rhodes et al. 2003, Massey et al. 2003, Refregier et
al. 2003 \\ $^{c}$ the survey will be complete in 2003\\ $^{d}$ a
survey of several 100 deg$^{2}$ is planned\\ $^{e}$ assuming the
availability of the optical camera\\}

\end{minipage}
\end{table*}

Radio surveys offer another interesting prospect. Ongoing efforts are
aimed at detecting cosmic shear with the FIRST radio survey (Chang \&
Refregier 2002, Kamionkowski et al. 1997, Refregier et al. 1998). The
future radio telescopes LOFAR (Low Frequency Array) and SKA
(Square Kilometer Array) will yield cosmic-shear measurements of
comparable sensitivity to the most ambitious optical surveys
(Schneider 1999). The advantages of radio surveys are that they cover
large solid angles, that the bright radio sources are at a higher
redshift, and that the PSF is fully predictable and reproducible.

\begin{figure}
\centerline{\psfig{figure=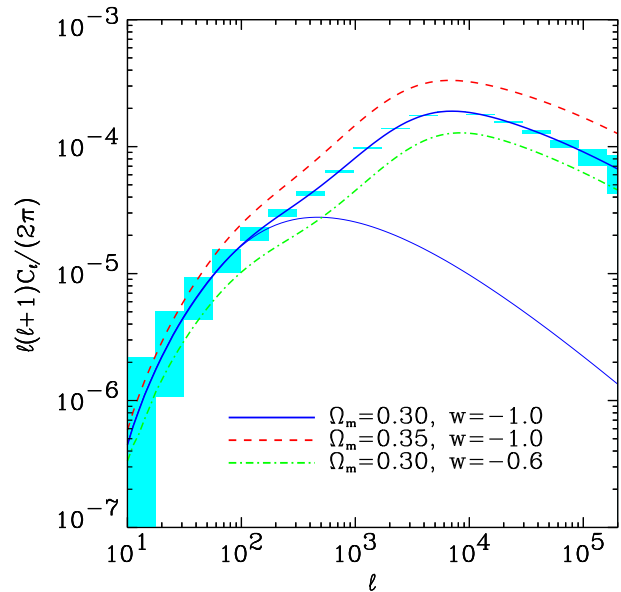,width=90mm}}
\caption{Prospects for the measurement of the weak-lensing power
spectrum with future weak-lensing surveys. The solid line represents
the lensing power spectrum of a $\Lambda$CDM model with
$\Omega_{m}=0.3$ (and an equation of state $w=-1$). The boxes
correspond to the band-averaged $1\sigma$ errors for the SNAP wide
survey (300 deg$^{2}$, 100 galaxies per arcmin$^{2}$ with a
root-mean-square intrinsic ellipticity dispersion of
$\sigma_{\epsilon}=0.31$). The precision is highest in the nonlinear
region [see the linear $\Lambda$CDM power spectrum ({\it dotted
line})] and will thus provide a test of the gravitational instability
paradigm. This model can easily be distinguished from models that
differ from it by a 17\% change in $\Omega_{m}$ ({\it dashed line}) or
by a 40\% change in the dark-energy equation of state $w$ ({\it
dot-dashed line}).}
\label{fig:cl_omw}
\end{figure}

These future surveys will provide very accurate measurements of
cosmological parameters through the measurement of the lensing power
spectrum and higher-order statistics.  Figure~\ref{fig:cl_omw} (see
color insert) shows, for instance, the excellent accuracy with which
the lensing power spectrum will be measured with the SNAP wide
survey. They will also allow us to test some of the foundations of the
standard cosmological model.  For example, the measurement of the
power spectrum on nonlinear scales at different redshift slices, and
of the hierarchy of high-order correlation functions, will yield a
direct test of the gravitational instability paradigm. The lensing
power spectrum can also be used to measure the equation of the state
of the dark energy $w$ and thus complement supernovae measurements in
the constraining of quintessence models (Benabed \& Bernardeau 2001;
Hui 1999; Huterer 2001; Hu 2001, 2002; Munshi \& Wang 2002; Weinberg
\& Kamionkowski 2002).  Figure~\ref{fig:cl_omw} shows, for instance,
that a change of 40\% in $w$ can easily be measured by SNAP (see
Refregier et al. 2003). Cosmic-shear measurements can also be used to
test general relativity (Uzan \& Bernardeau 2001).

Another promising approach to measure weak lensing is to use the
fluctuations of the CMB temperature as the background sources
(Bernardeau 1997, 1998; Cooray \& Kesden 2002; Hirata \& Seljak 2002;
Hu 2000; Seljak 1996; Seljak \& Zaldarriaga 1999; Zaldarriaga \&
Seljak 1998). Because these fluctuations are produced at a redshift of
approximately 1100, CMB lensing provides a probe of the
evolution of mass fluctuations at redshifts larger than those
probed by optical galaxies.  Lensing indeed produces distinct
non-Gaussian signatures that can be used to reconstruct the foreground
mass distribution and probe the growth of structures. This approach
will become feasible with the advent of future CMB missions
such as Planck Surveyor or ground-based instruments with
high angular resolutions.

\section{CHALLENGE: SYSTEMATIC EFFECTS}
\label{challenges} 
For the future surveys described above to yield their full potential,
a number of challenges must be met. Indeed, these surveys require the
measurement of shears on the order of 1\% with an accuracy better than
0.1\%; thus, they require a very tight control of systematic
effects. In the following, we review the main sources of systematics.

\subsection{PSF Anisotropy}
The most serious systematic for ground-based surveys is that
produced by the rather large PSF ellipticities ($\sim 10\%$)
observed by the different groups. The KSB method (see Section 4)
provides a correction of the PSF anisotropy by, at most, a factor
of approximately 10, which is insufficient for a precision of
0.1\% in shear. Although new shear measurement methods may improve
upon this (see Section 4 and below), the correction is
fundamentally limited by the finite surface density and
signal-to-noise ratio of the stars used to measure the PSF shape.
It is, therefore, necessary for the PSF ellipticity to be  reduced
in hardware as well as in software. This can be done by using
tighter constraints in the tracking and optical systems of the new
telescopes and instruments as well as with interactions between
telescope designers, engineers, and scientists.

\subsection{Shear Measurement Method}
Even if the PSF ellipticity were guaranteed to be isotropic, future
measurements will be limited by the precision of the shear measurement
methods. The KSB method is accurate to measure shears of approximately
1\% with only a 10\% accuracy (Bacon et al. 2001, Erben et
al. 2001). This will soon become an important limiting factor (see
Hirata \& Seljak 2003). Although new methods (see Section 4) promise
to improve upon this, more extensive simulations are required to
establish that the same accuracy can be reached for shears of 0.1\%.

\subsection{Redshift Distribution}
To convert cosmic-shear measurements into constraints on cosmological
parameters, the redshift distribution of the background galaxies must
be known (see Section 4). The uncertainty in the median galaxy
redshift is already one of the dominant contributions to the
uncertainty in the amplitude of the matter power spectrum from
cosmic shear. The determination of the galaxy redshift
distribution is made difficult by the depth of the cosmic-shear
surveys. In addition, the sample of galaxies is not simply magnitude
limited, but it is subject to complex selection cuts throughout the
shear measurement pipeline. Dedicated spectroscopic surveys and
photometric redshift studies (such as that by Brown et al. 2003)
are thus required to overcome this limitation.

\subsection{CCD Nonlinearities}
All the shear measurement methods rely on the assumption that the
instrumental response is linear. It is therefore important to test
whether CCD cameras do not have subtle pixel-to-pixel nonlinearities
that would induce biases in the shear measurements. The mean shear
offset and the shear gradient across the chip found by van Waerbeke et
al. (2001) could be due to this effect.

\subsection{Overlapping Isophotes}
Two neighboring galaxies in an image yield "peanut-shaped" isophotal
contours and thus appear to have aligned ellipticities. This
overlapping-ellipticities effect may produce a spurious ellipticity
correlation signal on small scales. van Waerbeke et al. (2000)
suggested that this effect may explain the excess shear signal, which
they measured on small scale and which disappeared when close pairs
were discarded.  Evidence for this effect was also found in the image
simulation by Bacon et al. (2001). More extensive and detailed
simulations would need to be performed to ascertain and calibrate this
effect.

\subsection{Intrinsic Correlations}
The measurement of the lensing shear relies on the assumptions that,
in the absence of lensing, the ellipticities of the galaxies are
uncorrelated. However, an intrinsic correlation of galaxy shapes could
exist owing to the coupling of the galaxy angular momentum or shape to
the tidal field or to galaxy interactions (see Heavens 2001 for a
review). Theoretical estimation of the size of this effect has been
performed using numerical simulations (Croft \& Metzler 2000; Heavens,
Refregier \& Heymans 2000; Jing 2002) and analytical methods (Catalan,
Kamionkowski \& Blandford 2001; Crittenden et al. 2001a,b; Lee \& Pen
2001; Mackey, White \& Kamionkowski 2001). Measurements of intrinsic
correlations have also been performed (Brown et al. 2000; Pen, Lee \&
Seljak 2000).  Although considerable uncertainty remains regarding the
amplitude of this effect, a consensus is arising that intrinsic
correlations are likely to be small for the deep current surveys with
$z_{m} \sim 1$, but may be dominant for shallower surveys with $z_{m}
\ls 0.2$ (see however the conflicting results of Jing 2002). Although
lensing distortions are coherent over a large redshift range,
intrinsic alignments are only significant for small physical
separations. Photometric redshifts can thus be used to separate and
reduce intrinsic correlations from cosmic-shear signals (Heymans \&
Heavens 2002; King \& Schneider 2002a,b).  Another approach is to
search for the $B$-type correlation signal produced by intrinsic
correlations (Crittenden et al. 2001b).
\label{intrinsic}

\subsection{Theoretical Uncertainties}
Most of the signal in cosmic-shear surveys arises from small scales
($\theta \ls 10'$) and thus from nonlinear structures (Jain \& Seljak
1997). The existing prescriptions for computing the nonlinear
corrections (Peacock \& Dodds 1997; Ma 1998) to the matter power
spectrum are only accurate to approximately 10\% and disagree at that
level with one another (see discussion in Huterer 2001). This
theoretical uncertainty will soon become one of the dominating errors
in the determination of cosmological parameters from cosmic-shear
surveys (see discussion in van Waerbeke et al. 2002). New
prescriptions based on more recent N-body simulations such as those by
Smith et al. (2003) will help improve the accuracy of the
predictions. The problem of the prediction of higher-order statistics
is even more difficult, but it is not as pressing given the large
uncertainties in the the measurements within current surveys (see
Bernardeau, Mellier, van Waerbeke 2002). The inclusion of second-order
terms in the weak-lensing approximation will eventually also be
required (Cooray 2002, Cooray \& Hu 2002).

These systematic effects must be controlled to match the statistical
accuracy of future surveys. Along with the studies suggested above,
further measurements with different colors and comparison of various
surveys in overlapping regions would help to control systematics and
to test data analysis pipelines.  On the theoretical side, larger
simulations coupled with advanced analytical techniques will be
required.

\section{CONCLUSIONS}
\label{conclusion}
Cosmic shear has emerged as a powerful method to measure the
large-scale structure in the universe. It can be thought of as the
measurement of background fluctuations in the space-time
metric. Although other methods rely on assumptions relating the
distribution of light to that of the mass, weak lensing is based on
``clean'' physics and can be directly compared to theory. The past
three years have yielded impressive observational progress, as the
first statistical detections and measurements of cosmic shear have
been achieved. In analogy with the CMB, cosmic shear has moved from
the COBE era to that of the first generation of anisotropy
experiments. However, the measurement of cosmic shear differs
from that of CMB anisotropies in several respects. First, the
fluctuations are on the order of $10^{-2}$ as opposed to $10^{-5}$,
making them easier to measure while retaining the validity of linear
calculations. Second, the cosmic-shear field is non-Gaussian and
therefore contains more information than that quantified by the power
spectrum.

Existing cosmic-shear measurements have started to yield significant
constraints on cosmological parameters. The measurement of the
amplitude of the matter power spectrum $\sigma_{8} \Omega_{m}^{0.5}$
from cosmic shear should soon replace that derived from the local
abundance of clusters. This latter technique is indeed limited by the
finite number of bright clusters and by systematic uncertainties in
the physics of clusters. With better statistics, the angular and
redshift dependence of the shear signal as well as with higher-order
moments of the convergence field will break the degeneracy between
$\sigma_{8}$ and $\Omega_{m}$ and yield constraints on further
parameters such as $\Omega_{\Lambda}$ and $\Gamma$.

The current measurements of cosmic shear are now primarily limited by
statistics. They will therefore be improved upon by a number of
upcoming and future instruments such as Megacam, VST, VISTA, LSST, and
Pan-STARRS from the ground, and HST/ACS and SNAP in space. For several
of these instruments, a weak-lensing survey has been listed as one of
the primary science drivers. These surveys will potentially yield
measurements of cosmological parameters that are comparable in
precision and complementary to those derived from the CMB. They will
also be able to address more far-reaching questions in cosmology by
measuring parameters beyond the standard model. For instance, they can
be used to provide a test of the gravitational instability paradigm, a
measure of the equation of state of the dark energy, and a test of
general relativity.

For these instruments to yield the full promise of cosmic shear, a
number of challenges have to be met. First, observationally,
systematic effects, such as the PSF anisotropy and CCD nonlinearities,
must be controlled and corrected for. From the theoretical point of
view, calculations of the nonlinear power spectrum, of high-order
statistics, and of the associated errors must be improved to meet the
precision of future measurements. The observational and theoretical
efforts required to overcome these difficulties are worthwhile given
the remarkable promise that cosmic shear offers to cosmology.

\noindent {\sc Acknowledgments} \noindent The author thanks
his collaborators David Bacon, Richard Massey, Tzu-Ching Chang, Jason
Rhodes, and Richard Ellis for numerous fruitful discussions. The
author thanks Bhuvnesh Jain, David Bacon, and Henk Hoekstra for their
permission to reproduce their figure. He also thanks Tony Tyson,
Herv\'{e} Aussel and Andy Taylor for precisions regarding the
parameters of future cosmic shear surveys.  He is also grateful to
Richard Ellis, Henk Hoekstra, Gary Bernstein, Yannick Mellier, Richard
Massey, and Ivan Valtchanov for useful comments on the manuscript.

\vglue .25in \centerline{The Annual Review of Astronomy and
Astrophysics is online at}
\centerline{http://astro.annualreviews.org}


\begin{thebibliography}{}
\bibitem{bab91} Babul, A Lee MH. 1991. {\it MNRAS} 250:407
\bibitem{bar01} Bartelmann M, King LJ, Schneider P. 2001. {\it Astron. Astrophys.} 378:361
\bibitem{bar99} Bartelmann M, Schneider P. 1999. astro-ph/9912508
\bibitem{bac01a} Bacon DJ, Refregier A, Clowe D, Ellis R. 2001. {\it MNRAS} 325:1065
\bibitem{bac00} Bacon DJ, Refregier A, Ellis R. 2000. {\it MNRAS} 318:625
\bibitem{bac02} Bacon DJ, Massey R, Refregier A, Ellis R. 2002. astro-ph/0203134
\bibitem{Ben01} Benabed K, Bernardeau F. 2001. {\it Phys. Rev. D} 64:083501
\bibitem{ben01} Bennet DP, Rhie SH. 2000. {\it GEST Home Page}.\hfill\\ {\tt http://bustard.phys.nd.edu/GEST. astro-ph/0011466}
\bibitem{ber97} Bernardeau F. 1997.  {\it Astron. Astrophys.} 324:15
\bibitem{ber98} Bernardeau F. 1998. {\it Astron. Astrophys.} 338:767
\bibitem{ber99} Bernardeau F. 1999. Theoretical and Observational Cosmology. {\it Proc. Cargese Summer Sch.}, {\it Cargese}, {\it France}, ed. M Lachieze-Rey. astro-ph/9901117
\bibitem{Ber02b} Bernardeau F, Mellier Y, van Waerbeke L. 2002. {\it Astron. Astrophys.} 389:L28
\bibitem{ber00} Bernardeau F, Valageas P. 2000. {\it Astron. Astrophys.} 364:1
\bibitem{Ber97} Bernardeau F, van Waerbeke L, Mellier Y. 1997. {\it Astron. Astrophys.} 322:1
\bibitem{Ber02a} Bernardeau F, van Waerbeke L, Mellier Y. 2003. 
  {\it Astron. Astrophys.} 397:405
\bibitem{ber01} Berstein GM, Jarvis M. 2001. astro-ph/0107431
\bibitem{b4} Blandford RD, Saust AB, Brainerd TG, Villumsen JV. 1991. {\it MNRAS} 241:600
\bibitem{bon95} Bonnet H, Mellier Y. 1995. {\it Astron. Astrophys.} 303:331
\bibitem{bor01} Borgani S, Rosati P, Tozzi P, Stanford SA, Eisenhardt PR, et al. 2001. {\it Ap. J.} 561:13
\bibitem{bou00} Boulade O, Charlot X, Abbon P, Aune S. Borgeaud P. et al. 2000. 
  {\it Proc. SPIE} 4008:657
  {\it Megacam Home Page.}\hfill\\ {\tt
  http://www-dapnia.cea.fr/Phys/Sap/Activites/Projets/Megacam/page.shtml}
\bibitem{bro00} Brown M, Taylor AN, Hambly N, Dye S. 2000. astro-ph/0009499
\bibitem{bro03} Brown M, Taylor AN, Bacon, D.J., Gray, M.E., Dye S.,
   Meisenheimer, K., Wolf, C. 2003. {\it MNRAS} 341:100
\bibitem{cat01} Catalan P, Kamionkowksi M, Blandford R. 2001. {\it MNRAS} 320:7
\bibitem{cha02} Chang T-C, Refregier A. 2002. {\it Ap. J.} 570:447
\bibitem{cont03} Contaldi, C, Hoekstra, H., Lewis, A. 2003.
  submitted to {\it Phys. Rev. Letters}. astro-ph/0302435
\bibitem{coo02} Cooray A. 2002. astro-ph/0206068
\bibitem{cooh01a} Cooray A, Hu W. 2001a. {\it Ap. J.} 548:7
\bibitem{cooh01b} Cooray A, Hu W. 2001b. {\it Ap. J.} 554:56
\bibitem{cooh02} Cooray A, Hu W. 2002. {\it Ap. J.} 574:19
\bibitem{coo00} Cooray A, Hu W, Miralda-Escud\'{e} J. 2000. {\it Ap. J.} 535:9
\bibitem{cook02} Cooray A, Kesden M. 2002. astro-ph/0204068
\bibitem{cri01a} Crittenden R, Natarajan P, Pen U, Theuns, T. 2001a. {\it Ap. J.} 559:552
\bibitem{cri01b} Crittenden R, Natarajan P, Pen U, Theuns T. 2001b. astro-ph/0012336
\bibitem{cro00} Croft RAC, Metzler CA. 2000. {\it Ap. J.} 545:561
\bibitem{eke98} Eke V, Cole S, Frenk C, Patrick Henry J. 1998. {\it MNRAS} 298:1145
\bibitem{erb01} Erben T, van Waerbeke L, Bertin E, Mellier  Y, Schneider P. 2001. {\it Astron. Astrophys.} 3667:17
\bibitem{b6} Fort B, Mellier Y. 1994. {\it Astron. Astrophys. Rev.} 5:239
\bibitem{gray02} Gray ME, Taylor AN, Meisenheimer K, Dye S, Wolf C, Thommes E. 2002. {\it Ap. J.} 568:141
\bibitem{gun67} Gunn JE. 1967. {\it Ap. J.} 150:737
\bibitem{ham00} Hamana T, Colombi S, Mellier Y. 2000. Cosmological Physics with Gravitational Lensing. {\it Proc. XXth Moriond Astrophys. Meet.}, {\it Les Arcs}, {\it France}, ed. J-P Kneib, Y Mellier, M Mon, J Tran Thanh Van. astro-ph/0009459
\bibitem{ham03} Hamana T, Miyazaki S, Shimasaku K, Furusawa H,
  Doi M, et al. 2002, submitted to {\it Ap. J.}, preprint
  astro-ph/0210450
\bibitem{Hae02} H\"{a}mmerle H, Miralles JM, Schneider P, Erben T, Fosbury RA. 
  2002. {\it Astron. Astrophys.} 385:743
\bibitem{hea01} Heavens AF. 2001. {\it Intrinsic Galaxy Alignments and Weak Gravitational Lensing}. Yale Worksh. Shapes Galaxies Haloes, May. astro-ph/0109063
\bibitem{hea00} Heavens A, Refregier A, Heymans C. 2000. {\it MNRAS} 319:649
\bibitem{hym02} Heymans C, Heavens A. 2002. astro-ph/0208220
\bibitem{hir02} Hirata C, Seljak U. 2002. astro-ph/0209489
\bibitem{hir03} Hirata C, Seljak U. 2003. to appear in {\it MNRAS}.
  astro-ph/0301054
\bibitem{hir01} Hiroyasu A, et al. 2001. {\it Subaru Home Page}. {\tt http://www.subaru.naoj.org/}
\bibitem{hoe98} Hoekstra H, Franx M, Kuijken K, Squires G. 1998. {\it Ap. J.} 504:636 
\bibitem{hoe01a} Hoekstra H, Franx M, Kuijken K, Carlberg RG, Yee HKC, et al.  2001. {\it Ap. J.} 548:5 
\bibitem{hoe01b} Hoekstra H, Yee HKC, Gladders, M. 2001. {\it Ap. J.} 558:11 
\bibitem{hoe02b} Hoekstra H, Yee HKC, Gladders M. 2002a. {\it Ap. J.} 
  577:595 
\bibitem{hoe02c} Hoekstra H, Yee HKC, Gladders M. 2002b. astro-ph/0205205 
\bibitem{hoe02a} Hoekstra H, Yee HKC, Gladders M, Felipe Barrientos L, Hall PB, Infante L. 2002a. {\it Ap. J.} 572:55 
\bibitem{hoe02d} Hoekstra H, van Waerbeke L, Gladders MD, Mellier Y, Yee HKC. 
  2002b. {\it Ap. J.} 577:604
\bibitem{hu99} Hu W. 1999. {\it Ap. J.} 522L:21
\bibitem{hu00} Hu W. 2000. {\it Phys. Rev. D} 63:3504
\bibitem{hu01} Hu W. 2001. astro-ph/010890
\bibitem{hu02} Hu W. 2002. astro-ph/0208093
\bibitem{huk02} Hu W, Keeton CR. 2002. astro-ph/0205412
\bibitem{hu+99} Hu W, Tegmark M. 1999. {\it Ap. J.} 514L:65
\bibitem{huw00} Hu W, White M. 2000. {\it Ap. J.} 554:67
\bibitem{hui00} Hui L. 1999. {\it Ap. J.} 519:9
\bibitem{hut01} Huterer D. 2001. astro-ph/0106399
\bibitem{b9} Jain B, Seljak U. 1997 {\it Ap. J.} 484:560.
\bibitem{jai00} Jain B, Seljak U, White S. 2000. {\it Ap. J.} 530:547
\bibitem{jai00b} Jain B, van Waerbeke L. 2000. {\it Ap. J.} 530:L1
\bibitem{jar90} Jaroszy\'{n}sky M, Park C, Paczy\'{n}sky B, Gott JR. 
  1990. {\it Ap. J.} 365:22
\bibitem{jar03} Jarvis, M., Bernstein, G.M., fisher, P., Smith, D.,
  Jain, B., Tyson, J.A., Wittman, D. 2002,{\it Ap. J.} 125:1014
\bibitem{jin02} Jing YP. 2002. {\it MNRAS} 335:L89
\bibitem{b12} Kaiser N. 1992. {\it Ap. J.} 388:272. 
\bibitem{b13} Kaiser N. 1998. {\it Ap. J.} 498:26. 
\bibitem{kai99} Kaiser N. 1999. {\it Weak Lensing by Galaxy Clusters}. Boston Lensing Meet. astro-ph/9912569 
\bibitem{kai00} Kaiser N. 2000. {\it Ap. J.} 537:555 
\bibitem{b14} Kaiser N, Squires G, Broadhurst T. 1995. {\it Ap. J.} 449:460 
\bibitem{kai+00b} Kaiser N, Tonry JL, Luppino GA. 2000. {\it PASP} 112:768.
  {\it Pan-STARRS homepage}
  {\tt http://pan-starrs.ifa.hawaii.edu/} 
\bibitem{kai+00} Kaiser N, Wilson G, Luppino GA. 2000. astro-ph/0003338 
\bibitem{kaiw99} Kaiser N, Wilson G, Luppino GA, Dahle H. 1999. astro-ph/99077229 
\bibitem{kai98} Kaiser N, Wilson G, Luppino G, Kofman L, Gioia I, et al.  1998. astro-ph/9809268 
\bibitem{kam97} Kamionkoski M, Babul A, Cress CM, Refregier A. 1997. {\it MNRAS} 301:1064. astro-ph/9712030
\bibitem{kin02a} King L, Schneider P. 2002a. astro-ph/0208256
\bibitem{kin02b} King L, Schneider P. 2002b. astro-ph/0209474
\bibitem{kri66} Kristian J, Sachs RK. 1966. {\it Ap. J.} 143:379
\bibitem{kri67} Kristian J. 1967. {\it Ap. J.} 147:864
\bibitem{kui99} Kuijken K. 1999. {\it Astron. Astrophys.} 352:355
\bibitem{kui02} Kuijken K, Bender R, Cappellaro E, Muschielok B, 
  Baruffolo A et al. 2002. {\it ESO Messenger} 110: 15; VST homepage 
  {\tt http://www.na.astro.it/vst/}
\bibitem{lee90} Lee MH, Paczy\'{n}sky B. 1990. {\it Ap. J.} 357:32
\bibitem{lee00} Lee J, Pen U. 2001. {\it Ap. J.} 532:5
\bibitem{lup97} Luppino GA, Kaiser N. 1997. {\it Ap. J.} 475:20
\bibitem{ma98} Ma C-P. 1998. {\it Ap. J.} 508:5
\bibitem{mac01} Mackey J, White M, Kamionkowksi M. 2001. astro-ph/0106364
\bibitem{mas03} Massey R., Rhodes J., Refregier A., Albert J., 
  Bacon D. et al. 2003. astro-ph/0304418
\bibitem{mao01} Maoli R, van Waerbeke L, Mellier Y, Schneider P, Jain B, et al.
  2001. {\it Astron. Astrophys.}, 368:766
\bibitem{mel99} Mellier Y. 1999. {\it Annu. Rev. Astron. Astrophys.} 37:127
\bibitem{mel02} Mellier Y, van Waerbeke L, Bertin E, Tereno I, Bernardeau F. 2002. astro-ph/0210091
\bibitem{mel01} Mellier Y, van Waerbeke L, Maoli R, Schneider P, Jain B, et al. 2001. Cosmic shear surveys.  {\it Deep Fields, Proc. Eur. South. Obs.}, {Oct.}, {\it Garching}, {Ger.} astro-ph/0101130
\bibitem{mel03} Mellier Y, van Waerbeke L, Radovich M, Bertin E, Dantel-Fort M.
   2000. {\it ESO Proceedings, Mining the Sky, Garching, July 2000, A.J. Banday et al eds. astro-ph/0012059}. {\it CFHTLS homepage
  {\tt http://cdsweb.u-strasbg.fr:2001/Science/CFHLS/}}
\bibitem{b17} Mould J, Blandford R, Villumsen J, Brainerd T, Smail I. 1994. 
  {\it MNRAS} 271:31
\bibitem{miy02} Miyazaki S, Hamana T, Shimasaku K, Furusawa H, Doi M, 
  et al. 2002. {\it Ap. J.} 580:97
\bibitem{mun00} Munshi D, Coles P. 2000. astro-ph/0003354
\bibitem{mun02} Munshi D, Coles P. 2002. astro-ph/0003481
\bibitem{mun+00} Munshi D, Jain B. 2000. {\it MNRAS} 318:109
\bibitem{mun01} Munshi D. Jain B. 2001. {\it MNRAS} 322:107
\bibitem{munw02} Munshi D, Wang Y. 2002. astro-ph/0206483
\bibitem{nar96} Narayan R, Bartelmann M. 1999. In {\it Formation of Structure in the Universe}, ed. A Dekel, JP Ostriker, p. 360. astro-ph/9606001
\bibitem{pad02} Padmanabhan N, Seljak U, Pen UL. 2002. {\it New Astronomy}
  8:581
\bibitem{pea96} Peacock J, Dodds SJ. 1997. {\it MNRAS} 280:L19
\bibitem{pen00} Pen U, Lee J, Seljak U. 2000. {\it Ap. J.} 543:L107
\bibitem{pen01} Pen U, van Waerbeke L, Mellier Y. 2001. astro-ph/0109182
\bibitem{per01} Perlmutter, et al. 2003. {\it SNAP Home Page}. {\tt http://snap.lbl.gov}
\bibitem{pie01} Pierpaoli E, Scott D, White M. 2001. {\it MNRAS} 325:77
\bibitem{pie02} Pierpaoli E, Scott D, White M. 2002. submitted to {\it MNRAS}.
  astro-ph/0210567
\bibitem{pre01} Premadi P, Martel H, Matzner R, Futamase T. 2001. {\it Ap. J. Suppl.} 135:7
\bibitem{ref01} Refregier A. 2003. {\it MNRAS} 338:35
\bibitem{refb01} Refregier A, Bacon DJ. 2003. {\it MNRAS} 338:48
\bibitem{rho99} Refregier A, Rhodes J, Groth E. 2002. {\it Ap. J.} 572:L131
\bibitem{b18} Refregier A, Brown ST, Kamionkowski M, Helfand DJ, Cress CM, 
  et al. 1998. Wide Field Surveys in Cosmology. {\it Proc. XIVth IAP Meet.}, 
  ed. Y Mellier, S Colombi. Paris. astro-ph/9810025
\bibitem{ref03} Refregier A, Massey M, Rhodes J, Ellis R, Albert J,
  et al. 2003. astro-ph/0304419 
\bibitem{rei01} Reiprich TH, B\"{o}hringer H. 2001. astro-ph/0111285
\bibitem{rho00} Rhodes J, Refregier A. Groth E. 2000. {\it Ap. J.} 536:79
\bibitem{rho01} Rhodes J, Refregier A, Groth E. 2001. {\it Ap. J.} 552:L85
\bibitem{rho03} Rhodes J., Refregier A., Massey R., Albert, J.,
  Bacon D., et al. 2003, astro-ph/0304417
\bibitem{sch95} Schneider P. 1995. {\it Proc. Laredo Adv. Summer Sch.}, Sept. astro-ph/9512047
\bibitem{sch98} Schneider P. 1998. {\it Ap. J.} 498:43
\bibitem{sc99} Schneider P. 1999. {\it Proc. Perspec. Radio Astron.}, {\it April}, {\it Amsterdam}. astro-ph/9907146
\bibitem{schl02} Schneider P, Lombardi M. 2002. astro-ph/0207454
\bibitem{b19} Schneider P, van Waerbeke L, Jain B, Kruse G. 1998a. {\it MNRAS} 296:873
\bibitem{sch02} Schneider P, van Waerbeke L, Kilbinger M, Mellier Y. 2002.
   {\it Astron. Astrophys} 396:1
\bibitem{sch88} Schneider P, Weiss A. 1988. {\it Ap. J.} 327:526
\bibitem{b20} Schneider P, van Waerbeke L, Mellier Y, Jain B, Seitz S, Fort B.
 1998b. {\it Astron. Astrophys.} 333:767
\bibitem{sel96} Seljak U. 1996. {\it Ap. J.} 463:1
\bibitem{sel01} Seljak U. 2001. astro-ph/0111362
\bibitem{sel99} Seljak U, Zaldarriaga M. 1999. {\it Phys. Rev. D} 60:43504
\bibitem{smi02} Smith RE, Peacock JA, Jenkins A, White SDM, Frenk CS, et al.
   2003. {\it MNRAS} 341:1311
\bibitem{sper03} Spergel, D., Verde, L., Peiris, H.V., Komatsu, E.,
  Nolta, M.R., et al. 2003, submitted to {\it Ap. J.}. astro-ph/0302209
\bibitem{ste96} Stebbins A. 1996. astro-ph/9609149
\bibitem{ste95} Stebbins A, McKay T, Frieman JA. 1995. {\it Proc. IAU Symposium 173}. astro-ph/9510012
\bibitem{tak02} Takada M. Jain B. 2002. astro-ph/0205055
\bibitem{tay01} Taylor A. 2001. astro-ph/0111605
\bibitem{wat01} Taylor A, Watts P. 2000. astro-ph/0010014
\bibitem{tay+03} Taylor A, et al. 2003. in preparation {\it VISTA Home Page.} {\tt http://www.vista.ac.uk}
\bibitem{tys02a} Tyson JA, Wittman D, Hennawi JF, Spergel DN. 2002a. 
 {\it Proc. 5th Int. UCLA Symp. Sources Detect. Dark Matter}, {Feb.}, {\it Marina del Rey}, ed. D Cline. astro-ph/0209632
\bibitem{tys02b} Tyson JA, \& the LSST collaboration 2002b. 
 {\it Proc. SPIE Int.Soc.Opt.Eng.}
 4836, 10-20. astro-ph/0302102.
 {\it LSST Home Page} {\tt http://lsst.org}
\bibitem{uza01} Uzan J-P, Bernardeau F. 2001. {\it Phys. Rev. D} 64:083004
\bibitem{val00} Valageas P. 2000. {\it Astron. Astrophys.} 356:771
\bibitem{val83} Valdes F, Jarvis JF, Tyson JA. 1983. {\it Ap. J.} 271:431
\bibitem{van98} van Waerbeke L. 1998. {\it Astron. Astrophys.} 334:1
\bibitem{van+98} van Waerbeke L, Bernardeau F, Mellier Y. 1999. {\it Astron. Astrophys.} 342:15
\bibitem{van00} van Waerbeke L, Mellier Y., Erben T., Cuillandre JC, 
  Bernardeau F, et al. 2000. {\it Astron. Astrophys.} 358:30 
\bibitem{van01a} van Waerbeke L, Mellier Y., Radovich M., Bertin E., 
  Dantel-Fort M. et al.  2001a. {\it Astron. Astrophys.} 374:757 
\bibitem{van01b} van Waerbeke L, Hamana T, Scoccimarro R, Colombi S, Bernardeau F. 2001b. {\it MNRAS} 322:918 
\bibitem{van02a} van Waerbeke L, Mellier Y, Pell\'{o} R, Pen U-L, McCracken HJ, Jain B. 2002. {\it Astron. Astrophys.} 393:369 
\bibitem{via99} Viana P, Liddle A. 1999. {\it MNRAS} 303:535
\bibitem{via01} Viana P, Nichol RC, Liddle A. 2001. astro-ph/0111394
\bibitem{vil95} Villumsen J. 1995. astro-ph/9507007
\bibitem{vil96} Villumsen J. 1996. {\it MNRAS} 281:369
\bibitem{van02b} van Waerbeke L, Mellier Y, Tereno I. 2002. astro-ph/0206245 
\bibitem{wam98} Wambsganss J, Cen R, Ostriker JP. 1998. {\it Ap. J.} 494:29
\bibitem{wein02} Weinberg NN, Kamionkowski M. 2002. astro-ph/0210134
\bibitem{whi99} White M, Hu W.  2000. {\it Ap. J.} 537:1
\bibitem{wil01} Wilson G, Kaiser N, Luppino GA. 2001. {\it Ap. J.} 556:601
\bibitem{wit00} Wittman DM, Tyson J, Kirkman D, Dell'Antonio I, Bernstein G. 2000. {\it Nature} 405:143
\bibitem{wit02} Wittman DM. 2002. {\it Dark Matter and Gravitational Lensing}, {\it LNP Top. Vol.}, ed. F Courbin, D Minniti. Springer-Verlag. astro-ph/0208063
\bibitem{wit+02} Wittman DM, Tyson JA, Dell'Antonio IP, Becker AC, 
  Margoniner VE, et al. 2002. {\it Proc. SPIE} 4836 v.2. astro-ph/0210118.
  {\it Deep Lens Survey web page} {\tt http://dls.bell-labs.com/}
\bibitem{zal02} Zaldarriaga M, Scoccimarro R. 2002. astro-ph/0208075
\bibitem{zal98} Zaldarriaga M, Seljak U. 1998. astro-ph/9810257


\end{thebibliography}
\end{document}